\documentclass[conference]{IEEEtran}
\usepackage{amsmath,amssymb,amsfonts}
\usepackage{algorithmic}
\usepackage{algorithm}
\usepackage{array}
\usepackage[caption=false,font=normalsize,labelfont=sf,textfont=sf]{subfig}
\usepackage{textcomp}
\usepackage{stfloats}
\usepackage{url}
\usepackage{verbatim}
\usepackage{graphicx}
\usepackage{caption}
\usepackage{subfig}  
\usepackage{graphicx}
\usepackage{titlesec}
\usepackage[hidelinks]{hyperref}
\usepackage{cite}
\usepackage{booktabs}
\usepackage{xcolor}
\usepackage{epsfig}
\usepackage{epstopdf}
\usepackage{acronym}
\graphicspath{{fig/}}
\usepackage{mathtools}
\usepackage{amsthm}
\usepackage{bm}
\usepackage{comment}
 \usepackage{setspace}
  \usepackage{graphicx}
  \usepackage{lipsum}
  \usepackage[absolute,overlay]{textpos}

\usepackage{ragged2e}

\newacro{ACDD}{Alamouti with cyclic delay diversity}
\newacro{URLLC}{ultra-reliable low-latency communications}
\newacro{3GPP}{third generation partnership project}
\newacro{PHY}{physical layer}
\newacro{MIMO}{multiple-input multiple-output}
\newacro{SIMO}{single-input multiple-output}
\newacro{MISO}{multiple-input single-output}
\newacro{SISO}{single-input single-output}
\newacro{MRC}{maximum-ratio combining}
\newacro{SNR}{signal-to-noise ratio}
\newacro{CP}{cyclic prefix}
\newacro{CDD}{cyclic delay diversity}
\newacro{FSC}{frequency-selective channel}
\newacro{STC}{space-time coding}
\newacro{FFT}{fast Fourier transform}
\newacro{LMMSE}{linear minimum mean-squared error}
\newacro{FER}{frame error rate}
\newacro{OFDM}{orthogonal frequency division multiplexing}
\newacro{OCDM}{orthogonal chirp division multiplexing}
\newacro{FSC}{frequency-selective channel}
\newacro{CSI}{channel state information}
\newacro{LMMSE-PIC}{linear minimum mean squared error with parallel interference cancellation}
\newacro{PFE}{perfect-feedback equalizer}
\newacro{FD}{full-duplex}
\newacro{PDP}{power delay profile}
\newacro{PDF}{probability density function}
\newacro{DFT}{discrete Fourier transform}
\newacro{SDFT}{sparse DFT}
\newacro{ICI}{inter-carrier interference}
\newacro{OTFS}{orthogonal time frequency space}
\newacro{AWGN}{additive white Gaussian noise}
\newacro{SWH}{sparse Walsh-Hadamard}
\newacro{LLR}{log-likelihood ratio}
\newacro{PMF}{probability mass function}
\newacro{CRC}{cyclic redundancy check}
\newacro{PAM}{pulse amplitude modulation}
\newacro{QAM}{quadrature amplitude modulation}
\newacro{FWHT}{fast Walsh-Hadamard transform}
\newacro{MAP}{maximum a-posteriori}
\newacro{SC}{single-carrier}
\newacro{ISI}{inter-symbol interference}
\newacro{ZP}{zero-padding}
\newacro{EVD}{eigenvalue decomposition}
\newacro{BCJR}{Bahl, Cocke, Jelinek, and Raviv}
\newacro{WHT}{Walsh-Hadamard transform}
\newacro{APP}{a-posteriori probability}
\newacro{SILE-EPIC}{self-iterated linear equalizer with expectation propagation}
\newacro{EP}{expectation propagation}
\newacro{i.i.d.}{independent and identically distributed}
\newacro{CWCU}{component wise conditionally unbiased}
\newacro{MSE}{mean squared error}
\newacro{EXIT}{extrinsic information transfer}
\newacro{MI}{mutual information}
\newacro{PAPR}{peak-to-average power ratio}
\newacro{DFT-s}{discrete Fourier transform-spread}
\newacro{AMP}{approximate message passing}
\newacro{GAMP}{generalized \ac{AMP}}
\newacro{VAMP}{vector \ac{AMP}}
\newacro{RSC}{recursive systematic convolutional}
\newacro{QPSK}{quadrature phase-shift keying}
\newacro{CFAR}{constant false alarm rate}
\newacro{PD}{probability of detection}
\newacro{PFA}{probability of false alarm}
\newacro{RV}{random variable}
\newacro{CDF}{cumulative distribution function}
\newacro{HD-ZP}{half-duplex ZP}
\newacro{FD-CP}{full-duplex ZP}
\newacro{DFRC}{dual-function radar communication}
\newacro{SINR}{signal-to-interference noise ratio}
\newacro{ISAC}{integrated sensing and communication}
\newacro{SI}{self-interference}
\newacro{RSI}{residual self-interference}
\newacro{ADC}{analog-to-digital converter}
\newacro{DAC}{digital-to-analog converter}
\newacro{ED}{energy-detection}
\newacro{IDFT}{inverse discrete Fourier Transform}
\newacro{SFFT}{symplectic finite Fourier transform }
\newacro{CRB}{Cram{\'{e}}r-Rao bound}
\newacro{ZC}{Zadoff-Chu}
\newacro{RMSE}{root mean square error}
\newacro{UW}{unique word}
\newacro{GFDM}{generalized frequency division multiplexing}
\newacro{RRC}{root-raised cosine}
\newacro{UB}{upper bound}
\newacro{CEF}{channel estimation field}
\newacro{TRX}{transceiver}
\newacro{IF}{intermediate frequency}
\newacro{RF}{radio frequency}
\newacro{FPGA}{field programmable gate arrays}
\newacro{SDR}{software-defined radio}
\newacro{UWB}{ultra wideband}
\newacro{PCB}{printed circuit board}
\newacro{SMA}{SubMiniature version A}
\newacro{MUSIC}{multiple signal classification}
\newacro{CIR}{channel impulse response}
\newacro{FR}{Frequency Range}
\newacro{mmWave}{millimeter wave}
\newacro{LoS}{line-of-sight}

\newcommand{\arm}{\mathrm{a}}
\newcommand{\brm}{\mathrm{b}}

\def\BibTeX{{\rm B\kern-.05em{\sc i\kern-.025em b}\kern-.08em
		T\kern-.1667em\lower.7ex\hbox{E}\kern-.125emX}}
	
\usepackage{soul,color}

\usepackage{tikz}
\usepackage{pgfplots}
\usetikzlibrary{shapes,arrows}
\usetikzlibrary{positioning,calc}
\usetikzlibrary{decorations.pathreplacing,calligraphy}
\usetikzlibrary{arrows.meta}
\usepackage{siunitx}

\tikzset{add/.style n args={4}{
		minimum width=3mm,
		path picture={
			\draw[black] 
			(path picture bounding box.south east) -- (path picture bounding box.north west)
			(path picture bounding box.south west) -- (path picture bounding box.north east);
			\node at ($(path picture bounding box.south)+(0,0.13)$)     {\tiny #1};
			\node at ($(path picture bounding box.west)+(0.13,0)$)      {\tiny #2};
			\node at ($(path picture bounding box.north)+(0,-0.13)$)        {\tiny #3};
			\node at ($(path picture bounding box.east)+(-0.13,0)$)     {\tiny #4};
		}
	}
}

\tikzset{add2/.style n args={4}{
		minimum width=1mm,
		path picture={
			\draw[black] 
			(path picture bounding box.south) -- (path picture bounding box.north)
			(path picture bounding box.west) -- (path picture bounding box.east);
			\node at ($(path picture bounding box.south)+(0,0.13)$)     {\tiny #1};
			\node at ($(path picture bounding box.west)+(0.13,0)$)      {\tiny #2};
			\node at ($(path picture bounding box.north)+(0,-0.13)$)        {\tiny #3};
			\node at ($(path picture bounding box.east)+(-0.13,0)$)     {\tiny #4};
		}
	}
}

\usepackage{pgfplots}
\usepgfplotslibrary{polar}
\usepackage{tikz}

\definecolor{applegreen}{rgb}{0.55, 0.71, 0.0}
\definecolor{awesome}{rgb}{1.0, 0.13, 0.32}
\definecolor{azure(colorwheel)}{rgb}{0.0, 0.5, 1.0}
\definecolor{darklavender}{rgb}{0.45, 0.31, 0.59}
\definecolor{cyan(process)}{rgb}{0.0, 0.72, 0.92}
\definecolor{brightmaroon}{rgb}{0.76, 0.13, 0.28}
\definecolor{ao(english)}{rgb}{0.0, 0.5, 0.0}
\definecolor{brightturquoise}{rgb}{0.03, 0.91, 0.87}
\definecolor{bondiblue}{rgb}{0.0, 0.58, 0.71}
\definecolor{atomictangerine}{rgb}{1.0, 0.6, 0.4}
\definecolor{classicrose}{rgb}{0.98, 0.8, 0.91}
\definecolor{copperrose}{rgb}{0.6, 0.4, 0.4}

\usepackage{etoolbox}

\begin{document}
\begin{textblock*}{20cm}(1cm, 0.25cm) 
\begin{flushleft}
Bomfin, R., Bazzi, A. Guo, H., Lee, H., Mezzavilla, M., Rangan, S., Choi, J., Chafii, M., ``An Experimental Multi-Band Channel Characterization in the Upper Mid-Band,'' \textit{Submitted to IEEE International Conference on Communications (ICC) 2025}, Montreal, Canada, Jun. 2025.
\end{flushleft}
\end{textblock*}

\title{An Experimental Multi-Band Channel Characterization in the Upper Mid-Band}

\author{Roberto Bomfin$^*$, Ahmad Bazzi$^{*\diamond}$, Hao Guo$^{\diamond}$$^{\S}$, Hyeongtaek Lee$^{\dagger}$, \\ Marco Mezzavilla$^\ddagger$, Sundeep Rangan$^{\diamond}$, Junil Choi$^{\dagger}$, and  Marwa Chafii$^{*\diamond}$ \\ 
	$^*$Engineering Division, New York University Abu Dhabi, UAE \\
	$^\diamond$NYU WIRELESS, NYU Tandon School of Engineering, New York, USA \\
    $^\S$Department of Electrical Engineering, Chalmers University of Technology, Gothenburg, Sweden\\
    $^\dagger$School of Electrical Engineering, KAIST, Republic of Korea \\
    $^\ddagger$Dipartimento di Elettronica, Informazione e Bioingegneria (DEIB), Politecnico di Milano, Milan, Italy\\
	Email: roberto.bomfin@nyu.edu, ahmad.bazzi@nyu.edu,  hg2891@nyu.edu, htlee8459@kaist.ac.kr, \\marco.mezzavilla@polimi.it, srangan@nyu.edu,   junil@kaist.ac.kr, marwa.chafii@nyu.edu   }

\maketitle

\begin{abstract}
The following paper provides a multi-band channel measurement analysis on the frequency range (FR)3.
This study focuses on the FR3 low frequencies \SI{6.5}{\GHz} and \SI{8.75}{\GHz} with a setup tailored to the context of integrated sensing and communication (ISAC), where the data are collected with and without the presence of a target.
A method based on multiple signal classification (MUSIC) is used to refine the delays of the channel impulse response estimates.
The results reveal that the channel at the lower frequency \SI{6.5}{\GHz} has additional distinguishable multipath components in the presence of the target, while the one associated with the higher frequency \SI{8.75}{\GHz} has more blockage. 
The set of results reported in this paper serves as a benchmark for future multi-band studies in the FR3 spectrum.
\end{abstract}


\section{Introduction}
%

The demand for new frequency bands such as \ac{FR}3 arises from the congestion in the sub-6 GHz spectrum and limitations of \ac{mmWave} frequencies. Sub-6 GHz bands, though widely used in cellular systems up to 4G, face severe spectral shortages. While \ac{mmWave} offers large bandwidths for multi-gigabit data rates in 5G, it suffers from limited transmission range and high blockage sensitivities, leading to inconsistent performance. \ac{FR}3, covering 7 to 24 GHz, presents a compelling alternative, balancing coverage and capacity, and is increasingly seen as a key resource for 5G-Advanced and 6G networks \cite{chafii2023twelve, kang2024cellular}.

\ac{FR}3 offers more spectrum than sub-6 GHz bands while retaining better propagation characteristics compared to \ac{mmWave} frequencies. This makes it ideal for cellular services that require a balance of bandwidth and coverage. However, there are existing services partially occupying the \ac{FR}3 bands, e.g., military radar, radio astronomy, and satellite communications, which makes spectral sharing crucial here \cite{kang2024cellular}. Regulatory bodies, such as the FCC, have started identifying parts of the \ac{FR}3 spectrum for cellular use, highlighting its growing importance in future spectrum allocations \cite{FCC20-443}.

Channel measurements in \ac{FR}3 are crucial for understanding its propagation characteristics and optimizing its use across various applications \cite{rappaport2024point,Shakya2025etal}. While recent measurement campaigns have gathered valuable data on, e.g., \ac{CIR}, there remains a significant gap in applying FR3 to target detection, particularly in challenging environments \cite{10694319}. The ability to use \ac{FR}3 frequencies for detecting objects, especially in scenarios where \ac{LoS} is weak or partially blocked, holds great potential for applications such as autonomous vehicles and radar-based sensing systems. However, the behavior of \ac{FR}3 frequencies in target detection is not yet fully understood, and existing models do not adequately capture the correlations across multiple frequencies that are necessary for accurate detection \cite{10693976}. Developing more adaptive models based on detailed multi-frequency measurements is essential for fully realizing the potential of \ac{FR}3 in these critical sensing applications.

In this paper, we conduct a multiple-antenna multi-band channel measurement campaign in the low FR3 region, in particular, at $\SI{6.5}{\GHz}$\footnote{While this frequency is at the edge of FR3, we avoid making this distinction explicitly for simplicity.} and $\SI{8.75}{\GHz}$.
In our system, we make use of the newly developed Pi-Radio \ac{RF} frontend \cite{10694319}, which is capable of up-converting signals spanning the whole FR3 range.
Moreover, with the recent studies of \ac{ISAC} in the literature \cite{LiuJSAC,Bomfin_UW,10018908,10373185}, we focus our investigation on comparing the multi-frequency channel measurements in the presence of a target.
In this initial set of results reported in this paper, we investigate how the impulse response changes from frequency to frequency and provide insights for ISAC.
In terms of processing techniques, we utilize a similar method to \cite{BomfinJCAS} where a \ac{MUSIC}-type algorithm is applied on discrete-time channel impulse response to provide a refined delay estimation of the multipath.
The novelty of this work consists of performing the multi-band channel measurement at the new FR3 spectrum and developing a methodology for the data analysis.
The results reveal that the channel at the lower frequency \SI{6.5}{\GHz} has additional multipath components in the presence of the target, while the higher frequency \SI{8.75}{\GHz} has more blockage.
This demonstrates a difference in how the target can alter the background channel (or clutter) depending on the frequency.
The methodology and results presented in this paper will serve as a benchmark for future measurements in the FR3 regions.
The contributions of this paper are summarized as

\begin{itemize}
    \item \textbf{Multi-band multi-antenna measurement in the new FR3 spectrum,} at $6.5 \operatorname{GHz}$ and $8.75 \operatorname{GHz}$, utilizing the Pi-Radio \ac{RF} frontend, which supports \textit{the entire FR3 range}. Our compact platform allows us to conduct FR3 measurements over multi-frequency channel measurements in the presence/absence of a target.
    \item \textbf{Methodology for ISAC channel characterization}, so as to highlight the prominent multipath components and cluster them into delay groups that are close to each other. It operates on a frame-by-frame basis, utilizing frequency-domain subcarrier smoothing to de-correlate the multipath components. Subsequently, a $K$-means clustering algorithm is employed, using a silhouette-based criterion to determine the number of delay clusters.
    \item \textbf{Insights for ISAC and benchmark for future multi-band measurement campaigns.} Insights reveal distinct characteristics between different FR3 frequencies. For instance, the presence of the target can introduce new paths or block previously existing ones differently for each frequency. This effect is further analyzed using positive and negative regions, which highlight the correlation of the target with the environment it is placed in. 
\end{itemize}



\section{System Model}\label{sec:methodology}
\subsection{Transmit Sequences}\label{subsec:transmit_sequences}
Our system is $2 \times 1$ \ac{MISO}, where two orthogonal sequences are transmitted in each antenna.
The discrete-frequency domain signal associated with the first antenna, subscript $\arm$, is $\mathbf{X}_{\rm a}\in \mathcal{S}_{\rm QPSK}^{N_{\rm on}}$, where $\mathcal{S}_{\rm QPSK}$ is the set of \ac{QPSK} symbols, and $N_{\rm on}$ is the number of allocated subcarriers centered at the middle of the spectrum.
We generate an orthogonal signal for the second antenna, subscript $\brm$, 
as $\mathbf{X}_{\rm b}[k] = \mathbf{X}_{\rm a}[k] e^{j2 \pi k/2} \,\, \forall k \in [-N/2+1,N/2]$ so that its discrete-time domain counterpart is $\mathbf{x}_{\rm b} = \mathbf{x}_{\rm a} \circledast \boldsymbol{\delta}_{N/2}$, where $\boldsymbol{\delta}_{i}[n]$ is the Kronecker delta vector with $1$ for $i=n$ and $0$ otherwise, and $\circledast$ represents the circular convolution.
The quantity $N$ represents the \ac{FFT} size.
Assuming a limited delay spread smaller than $N/2$ samples, the above property enables an easy MISO channel estimation by windowing the discrete-time domain channel estimate, as shown in \cite{Bomfin_CHest}.
The transmit signal at antenna $\arm$ is
\begin{equation}
    \tilde{x}_{\arm}(t) = \sum\nolimits_{n=-\infty}^{\infty} \tilde{\mathbf{x}}_{\rm a}[n] g_{\rm t}(t-n/B),
\end{equation}
where $g_{\rm t}(t)$ is the band-limited transmit filter with bandwidth $B$, and $\tilde{\mathbf{x}}_{\rm a} = [\mathbf{x}_{\rm a}^{\rm T} \,\, \mathbf{x}_{\rm a}^{\rm T} \cdots]$ is a concatenation of $\mathbf{x}_{\rm a}$ to describe its constant transmission.
Lastly, the input for antenna $\brm$ is obtained analogously.

\subsection{Wireless Channel Model} 
In this work, we explore transmissions in two different frequencies, namely, $f_{\rm c} = \SI{6.5}{\GHz}$ and $\SI{8.75}{\GHz}$.
The small-scale parameters such as the amplitude of each path, $\alpha_{\arm,f_{\rm c},l}$, are considered to be frequency-dependent due to the different reflection coefficients of objects.
The delay per path, $\tau_{l,\arm}$, and the total number of delays, ${L_\arm}$, are considered to be the same for both frequencies since they depend on the environmental geometry.
With the above considerations, the channel impulse response between the antenna $\arm$ and the receiver is defined as
\begin{equation}
    h_{\arm,f_{\rm c}}(t) = \sum\nolimits_{l=0}^{L_\arm-1} \alpha_{\arm,f_{\rm c},l}\delta(t-\tau_{l,\arm}),
\end{equation}
which is assumed to be time-invariant.
The received signal for $2\times 1$ MIMO system is given by the convolution between the transmitted signal and channel
\begin{equation}\label{eq:y_fc}
    \tilde{y}_{f_{\rm c}}(t) = \sum_{l=0}^{L_\arm-1} \alpha_{\arm,f_{\rm c},l}\tilde{x}_\arm(t-\tau_{l,\arm})+\sum_{l=0}^{L_\brm-1} \alpha_{\brm,f_{\rm c},l}\tilde{x}_\brm(t-\tau_{l,\brm}) + w(t),
\end{equation}
where $w(t)$ is the \ac{AWGN}.
\subsection{Channel Estimation}
Let ${y}_{f_{\rm c}}(t) = \tilde{y}_{f_{\rm c}}(t) * g_{\rm r}(t)$ be the filtered signal at the receiver, where $\mathbf{y}_{f_{\rm c}}[n] = y_{f_{\rm c}}(n/B)$ for $n \in [0,N-1]$ is the sampled received signal and $\mathbf{Y}_{f_{\rm c}}$ its DFT.
The discrete-frequency channel is obtained as ${\mathbf{R}}_{\arm,f_{\rm c}}[k] = \mathbf{Y}_{f_{\rm c}}[k]/\mathbf{X}_\arm[k]$,
%
%
which contains the channels $h_{\arm,f_{\rm c}}$ and $h_{\brm,f_{\rm c}}$ due to \eqref{eq:y_fc}.
The discrete-time channel impulse response for the antennas $\arm$ and $\brm$ are obtained by filtering the discrete-time signal ${{\mathbf{r}}_{\arm,f_{\rm c}} = {\rm IDFT}({\mathbf{R}}_{\arm,f_{\rm c}})}$ as \cite{Bomfin_CHest}
\begin{equation}\label{eq:h_hat}
    \begin{split}
        &\hat{\mathbf{h}}_{\arm,f_{\rm c}}[n] = \mathbf{r}_{\arm,f_{\rm c}}[n], \,\,\,\, 0\leq n < N/4 \\
        &\hat{\mathbf{h}}_{\brm,f_{\rm c}}[n] = (\mathbf{r}_{\arm,f_{\rm c}} \circledast \boldsymbol{\delta}_{-N/2})[n], \,\,\,\, 0\leq n < N/4,
    \end{split}
\end{equation}
where $\hat{\mathbf{h}}_{\arm,f_{\rm c}}[n] = \hat{\mathbf{h}}_{\brm,f_{\rm c}}[n] = 0$ for $n \geq N/4$.
This operation assumes that the first delays $\tau_{0,\arm}\approx \tau_{0,\brm} < 1/B$, which can be forced by appropriately shifting ${\mathbf{r}}_{\arm,f_{\rm c}}$ without problem since we are not interested in the absolute delays.
%
Lastly, the estimates of \eqref{eq:h_hat} in the frequency domain $\hat{\mathbf{H}}_{\arm,f_{\rm c}}, \hat{\mathbf{H}}_{\brm,f_{\rm c}} \in \mathbf{C}^{N_{\rm on}}$ are defined such that they contain only the allocated carriers. 

\section{Data Processing Methods}
\subsection{$2^{\rm{nd}}$-order Elbow Method \& MUSIC-PDP}
\label{sec-order-elbow-smooth}
In general, the covariance matrix associated with the parameters to be estimated plays a crucial role in numerous algorithms especially in array signal processing \cite{7952779}. 
Notice that we decompose the $ \mathbb{C}^{N_{\rm on} \times N_{\rm on}}$ covariance matrix of the channel estimates as
\begin{equation}\label{eq:C}
    \begin{split}
    \mathbf{C}_{\arm,f_{\rm c}}
    & =
    \mathbb{E}
    \left[
    \hat{\mathbf{H}}_{\arm,f_{\rm c}}
    \hat{\mathbf{H}}_{\arm,f_{\rm c}}^{\rm H}
    \right] = \mathbf{B}_{\arm}
     \boldsymbol{\alpha}_{\arm,f_{\rm c}}
     \boldsymbol{\alpha}_{\arm,f_{\rm c}}^H
     \mathbf{B}_{\arm}^H
     +
     \mathbf{W}_{\arm,f_{\rm c}},
    \end{split}
\end{equation}
where $\boldsymbol{\alpha}_{\arm,f_{\rm c}}\in \mathbb{C}^{L_\arm}$ is obtained by stacking up all elements of $\alpha_{\arm,f_{\rm c},l}$ in vectorized form.
Moreover, $\mathbf{B}_{\arm}$ is an $N_{\rm on} \times L_\arm$ matrix containing the path delays each parameterized within a column of $\mathbf{B}_{\arm}$.
Finally, $\mathbf{W}_{\arm,f_{\rm c}}$ is the covariance matrix of the noise process at carrier $f_{\rm c}$ and antenna $\arm$.
An analogous expression for antenna $\brm$.
It is straightforward to see that the signal subspace embedded within $\mathbf{C}_{\arm,f_{\rm c}}$ is $\operatorname{rank}$-$1$, where the underlying reason being that multipath channels are known to give rise to coherent signals \cite{5692980}, i.e. each path is nothing but a \textit{scaled and delayed} version of another.
This, in turn, implies that eigenmode analysis techniques do not directly reflect the total number of multipath components contained within a wireless propagation channel.

To remedy the rank-deficiency, we apply some well-known smoothing techniques across frequency domain \cite{7472290}.
Specifically, a space-frequency smoothing technique decorrelates the multipath components, hence restoring the rank of the signal subspace for eigenmode analysis purposes.
The outcome of the smoothing process is a smaller, but smoothed, covariance matrix version of \eqref{eq:C}, expressed as
$
    \tilde{\mathbf{C}}_{\arm,f_{\rm c}}
    =
     \tilde{\mathbf{B}}_{\arm}\boldsymbol{\alpha}_{\arm,f_{\rm c}}
     \boldsymbol{\alpha}_{\arm,f_{\rm c}}^H
     \tilde{\mathbf{B}}_{\arm}^H
     +
     \tilde{\mathbf{W}}_{\arm,f_{\rm c}} \in \mathbb{C}^{\tilde{N} \times \tilde{N}}
$
where now $\tilde{\mathbf{B}}_{\arm}$ is an $\tilde{N} \times L_\arm$ with $\tilde{N}<N_{\rm on}$.
After smoothing, a principle component analysis study is done via \ac{EVD}, which is performed on $\tilde{\mathbf{C}}_{\arm,f_{\rm c}}$ for model order selection. For this purpose, the eigenvalues are sorted in a vector $\mathbf{\lambda}_{\arm,f_{\rm c}}$, whereby an elbow-type method identifies an optimal index in $\mathbf{\lambda}_{\arm,f_{\rm c}}$ to determine the number of multipath components. 
The elbow method is based on the second-order statistics of the eigenvalues \cite{jolliffe2002principal}.
This optimal index, i.e. the \textit{elbow} point, should appear at $L_\arm$ eigenvalues, suggesting that the majority of the explained variability in the data can be captured within $L_\arm$ dimensions, with minimal loss of information.
After estimating $L_\arm$, denoted as $\widehat{L}_\arm$, we can now apply state-of-the-art parameter estimation methods, where we have chosen \ac{MUSIC} due to its ease of implementation.
The frequency-based smoothed \ac{MUSIC}-\ac{PDP} spectrum is formed as
\begin{equation}
\label{eq:MUSIC-PDP}
    P_{\arm,f_{\rm c}}(\tau)
    =
    \Vert \mathbf{b}^H(\tau) \mathbf{U}_{\arm,f_{\rm c}}^{\rm{noise}}  \Vert^{-2},
\end{equation}
where $\mathbf{U}_{\arm,f_{\rm c}}^{\rm{noise}}$ corresponds to the noise subspace obtained from $ \mathbf{C}_{\arm,f_{\rm c}}$.
It is worth noting that a \ac{MUSIC}-\ac{PDP} is much more robust to process compared to a classical \ac{PDP} spectrum because filters out outliers and tests for paths in the spectrum under maximum orthogonality, as per its criterion in \eqref{eq:MUSIC-PDP}.
After delay estimation, the path gains can be obtained by a simple inversion criterion.

\subsection{$K-$Means Clustering of multipath delays}
The $K-$Means is a commonly known and widely used clustering mechanism that aims at partitioning a given dataset into $K$ clusters. 
In our problem, it allows us to group the relative delays, relative to the first arriving path, clusters that are somehow close to one another.
For each symbol, the peaks of the MUSIC-PDP are collected and stored, as explained in Section \ref{sec-order-elbow-smooth}. 
These delay estimates are passed to a $K-$means clustering algorithm, that groups these delay estimates into groups.
The $K$-means clustering minimizes the sum of squared distances between the delay estimates and the clustered delay centroids, i.e. the objective would be to minimize the overall \textit{within-cluster sum of squares} clustering cost as
\begin{equation*}
    \lbrace \mathcal{T}_1^{\rm{opt}} \ldots \mathcal{T}_K^{\rm{opt}} \rbrace=\arg \min _{\mathcal{T}_1 \ldots \mathcal{T}_K} \sum\nolimits_{i=1}^K \sum\nolimits_{\tau_j \in \mathcal{T}_i}\left\|\tau_j-c_i\right\|^2,
\end{equation*}
where for sake of simplicity we denote all the collected delay estimates as $\tau_j$ and $c_i$ represents the clustered delay centroid of the $i^{th}$ cluster.
Due to lack of space, we omit the details of the $K-$means clustering algorithm. 
Moreover, we adopt a silhouette-type method, which reports how distinguishable the delay clusters $ \mathcal{T}_1^{\rm{opt}} \ldots \mathcal{T}_K^{\rm{opt}}$ are separated in order to determine the optimal number of delay clusters.
In short, the optimal number of clusters is the one that maximizes the so-called average silhouette score, which is the average of all silhouette coefficients, which for a given delay estimate is computed as $\xi\left(\tau_i\right)=\frac{\phi\left(\tau_i\right)-\Phi\left(\tau_i\right)}{\max \left\{\Phi\left(\tau_i\right), \phi\left(\tau_i\right)\right\}}$, where $\Phi\left(\tau\right)$ is the average distance from $\tau$ to all other points in the same cluster, where $\phi(\tau)$ is minimum average distance from $\tau$ to all points in neighboring delay clusters.

\section{Experimental Setup}\label{sec:experimental_setup}

\subsection{Hardware}
To implement our experiment in FR3, we propose and verify a novel \ac{SDR} platform for FR3, which can be configured as a transmitter (TX) or receiver (RX).
We design a system with $2$ transmit and $2$ receive chains.
The basic components of the proposed upper mid-band \ac{SDR} platform can be summarized as follows:
\begin{itemize}
    \item \textbf{Digital baseband board - RFSoC ZCU111}: 
    This evaluation Kit features the Zynq UltraScale+ XCZU28DR, integrating eight 12-bit \acp{ADC} and eight 14-bit \acp{DAC} with maximum sample rates of 4.096 Gsps and 6.554 Gsps, respectively. In our system, we utilize a Matlab interface to transfer baseband samples to and from the RFSoC.
    \item \textbf{\ac{RF} \ac{TRX} board - Pi-Radio board}: The Pi-Radio \ac{TRX}  board performs the up and down-conversion to 2 transmit and 2 receive \ac{IF} streams. This two-stage up- and down-conversion is used to meet the out-of-band and adjacent channel rejection ratio requirements and strict spectral masks. More details of the board can be found at \cite{10694319}.
	\item \textbf{Antenna - Kyocera patch antenna}: The Kyocera 1005193 is a \ac{UWB} \ac{PCB} antenna designed for operation over the 6.0–8.5 GHz frequency range\footnote{We have experimentally verified that these antennas can be used at $\SI{8.75}{\GHz}$.}. It offers a peak gain of 6.0 dBi, and has compact dimensions of 18.6 x 20.6 x 0.8 mm with easy integration with the Pi-Radio board.
\end{itemize}

\subsection{Environment}
\begin{figure}[!t]
\centering
\includegraphics[width=3.5in]{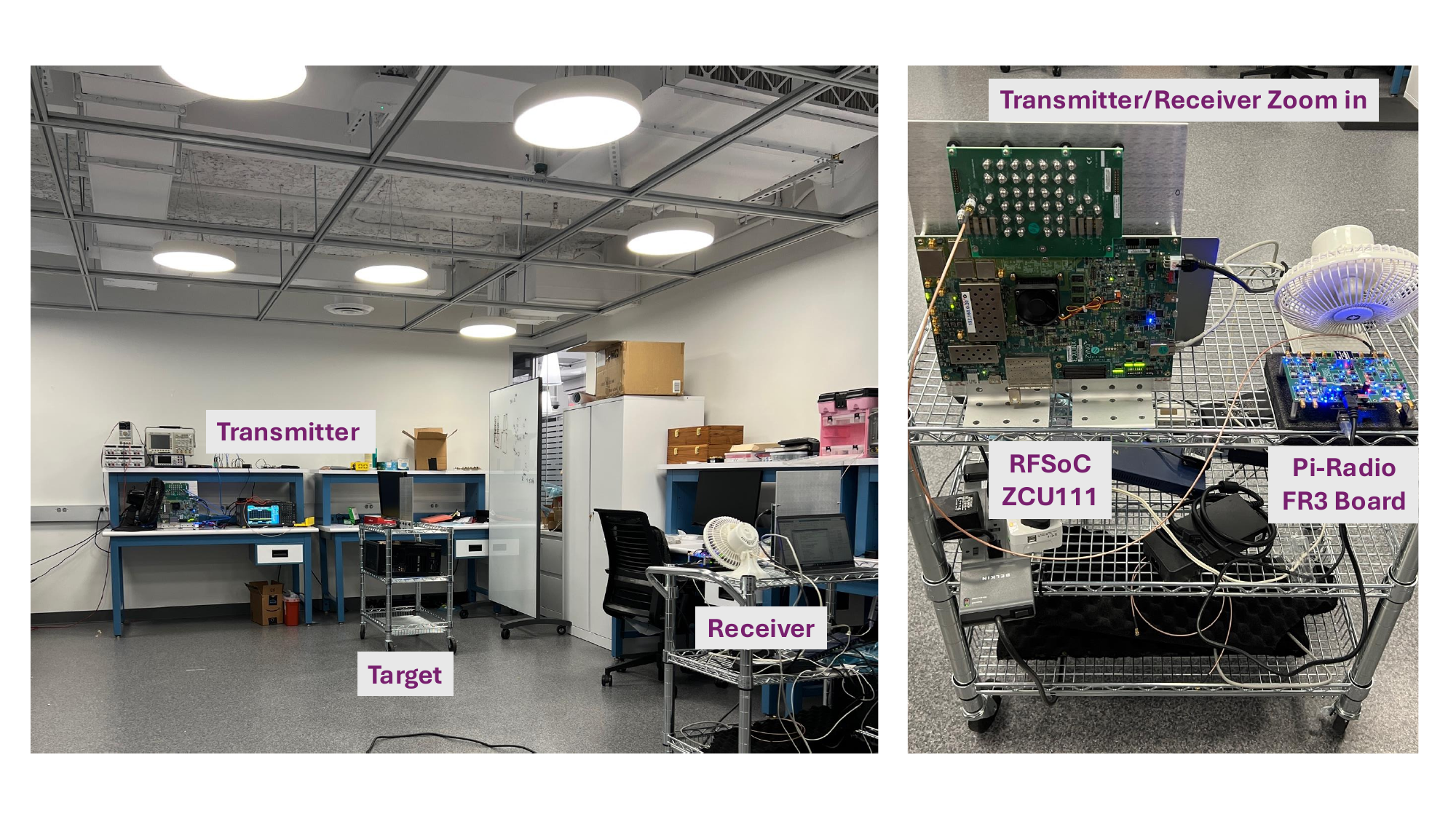}
\caption{Measurement environment at NYU Wireless, Brooklyn, NY, USA (https://wireless.engineering.nyu.edu/).}
\label{fig_lab}
\end{figure}

\begin{figure}
    \centering
    \subfloat[Locations]{
    \includegraphics[width=0.75\linewidth]{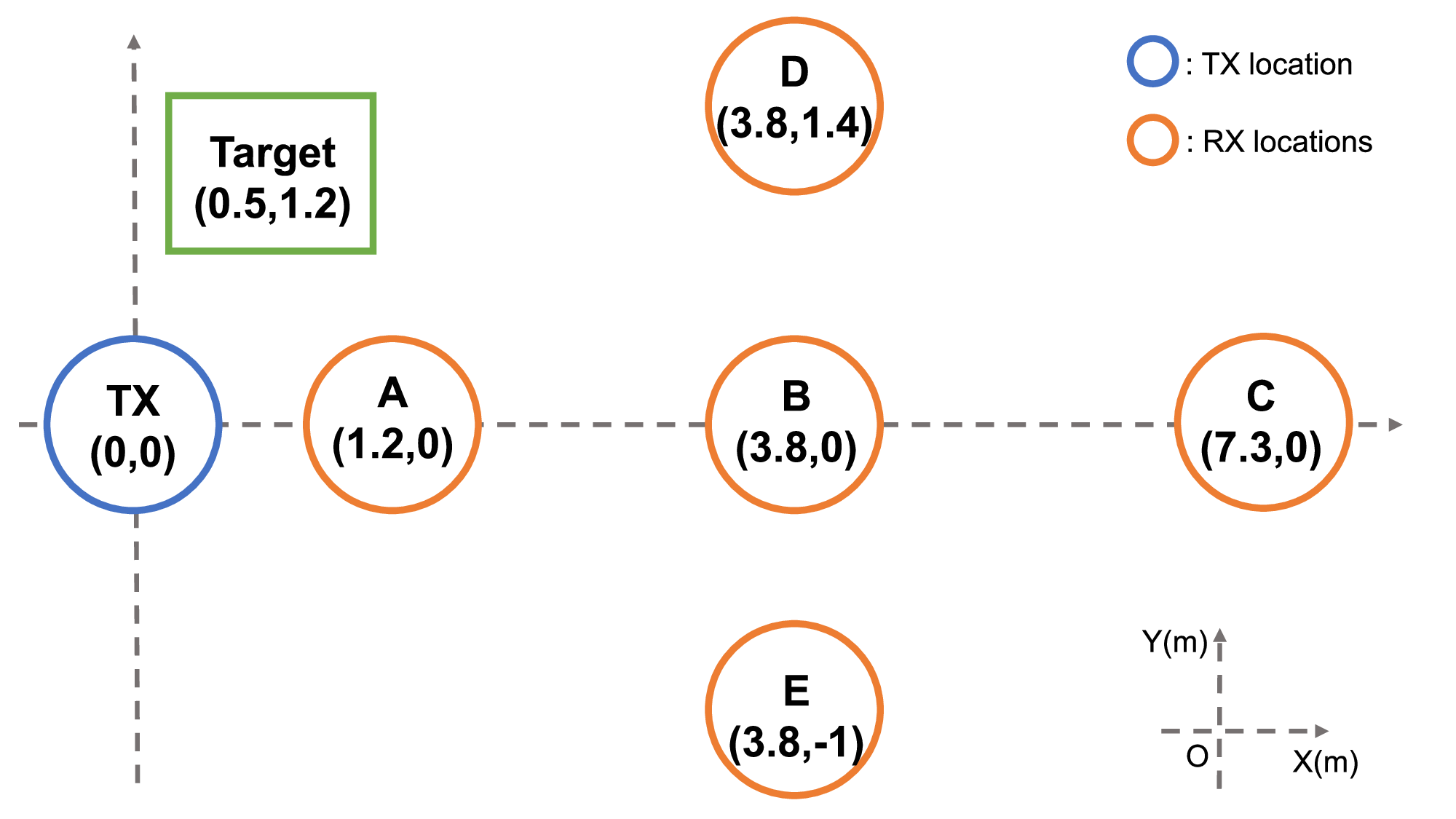}
    \label{fig_locations}
    }
    \hfill
    \subfloat[Antenna orientations]{
    \includegraphics[width=0.75\linewidth]{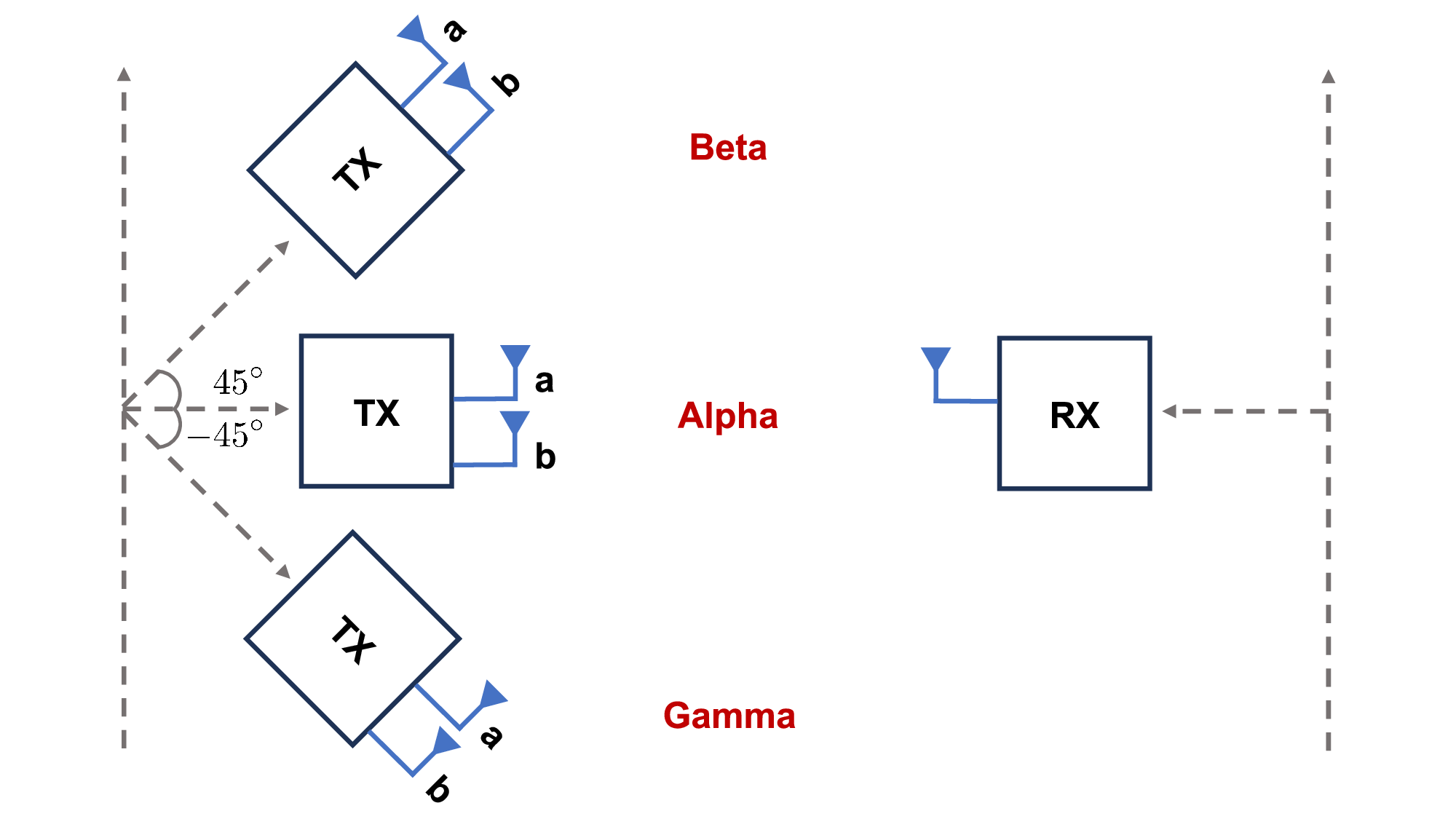}
    \label{fig_orientations}
    }
    \caption{Locations and antenna orientations setup.}    
\end{figure}

Our measurement campaign was performed in an indoor laboratory of NYU WIRELESS as shown in Fig. \ref{fig_lab} on the next page.
To have diversity in the data, we set the various locations of the TX and RX.
In addition, the measurements were conducted with and without the presence of a metallic structure denoted as the target.
The locations of the TX, RX, and target are depicted in Fig.~\ref{fig_locations}.
With the fixed locations for the TX and target, we considered five different RX locations, i.e., A to E, by setting the TX location as the origin, and the relative xy positions are represented. In addition, for each RX location, three different antenna orientations, i.e., Alpha, Beta, and Gamma, are assumed on the TX side as in Fig. \ref{fig_orientations}. 

\subsection{System Parametrization}
Regarding the system parameterization, we have used a number of $N_{\rm on} = 521$ subcarriers centered at $f_{\rm c}$, where the FFT size is set to $K = 1024$.
The sampling frequency is $B = 983.04 \, {\rm Hz}$ so that the bandwidth occupied by the transmitted signal is approximately $\SI{500}{\MHz}$.
For each channel measurement, a window of 1024 samples is averaged 100 times at the RFSoC ZCU111 to increase the SNR prior to channel estimation.
Then, 100 channel estimates are collected at the host computer for offline processing.

\section{Experimental Results}\label{sec:experimental_results}

\begin{figure}
    \centering
    \includegraphics[width=0.8\linewidth]{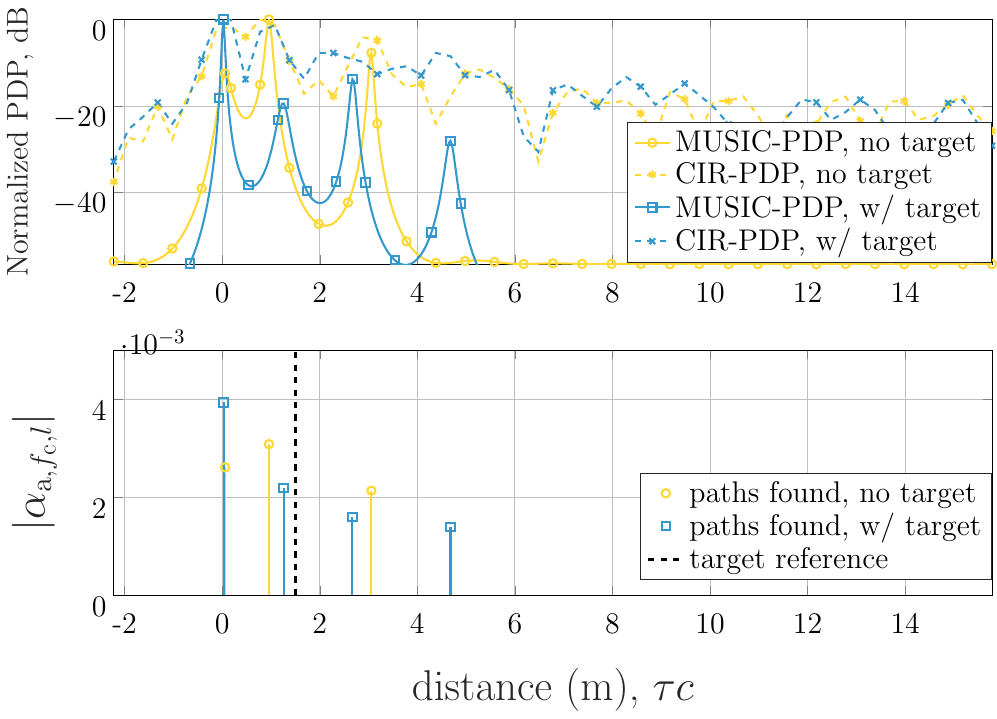}
    \caption{\textbf{Upper part}: The smoothed MUSIC-PDP vs a classical CIR-PDP. \textbf{Lower part}: The estimated path gains per path delay. The vertical dashed line corresponds to the target location only when a target of interest is present, and $\tau c$ is the additional distance traveled by the path in relation to the line-of-sight.}
    \label{fig:PDP_all}
\end{figure}

\begin{figure*}[h]
    \centering
    \begin{tikzpicture}
        \node[inner sep=0] at (0,0) {\includegraphics[width=0.27\textwidth]{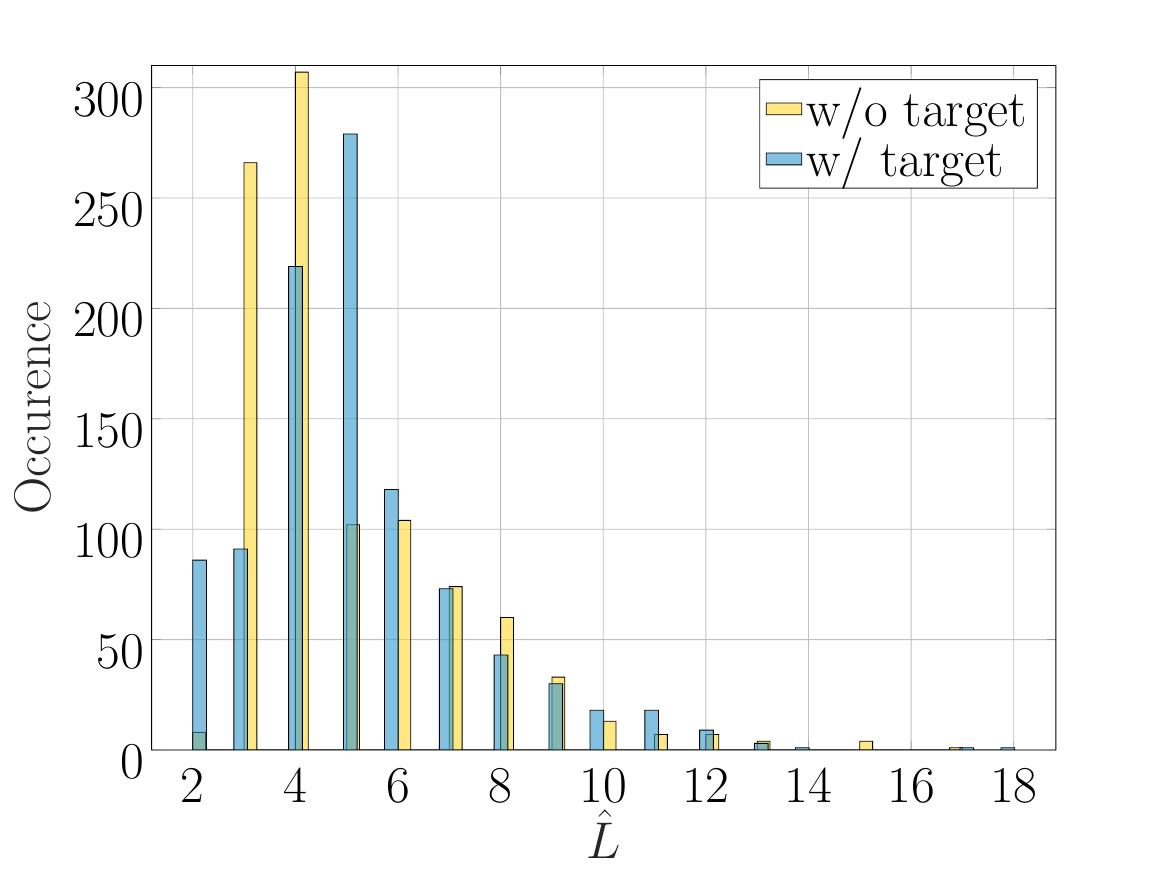}};
        \node[draw, fill=white, align=center] at (0.7,0.5) {\small Orientation: Beta};  
    \end{tikzpicture}
        \begin{tikzpicture}
        \node[inner sep=0] at (0,0) {\includegraphics[width=0.25\textwidth]{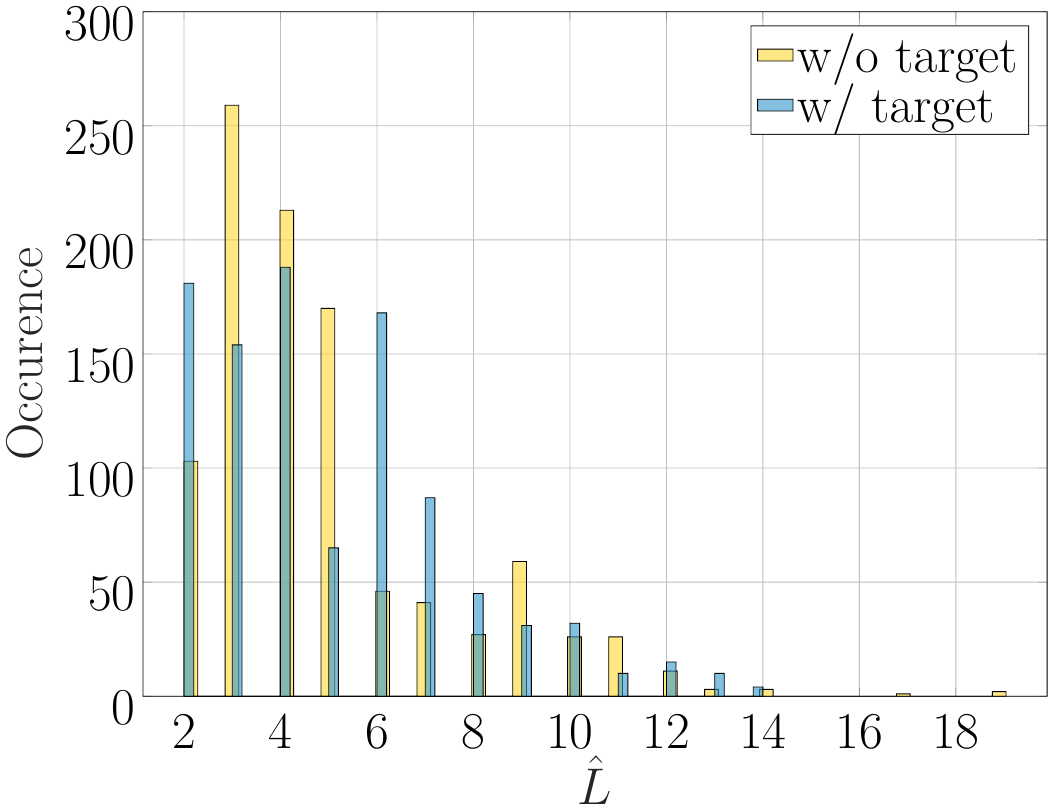}};
        \node[draw, fill=white, align=center] at (0.7,0.5) {\small Orientation: Alpha};
    \end{tikzpicture}
        \begin{tikzpicture}
        \node[inner sep=0] at (0,0) {\includegraphics[width=0.27\textwidth]{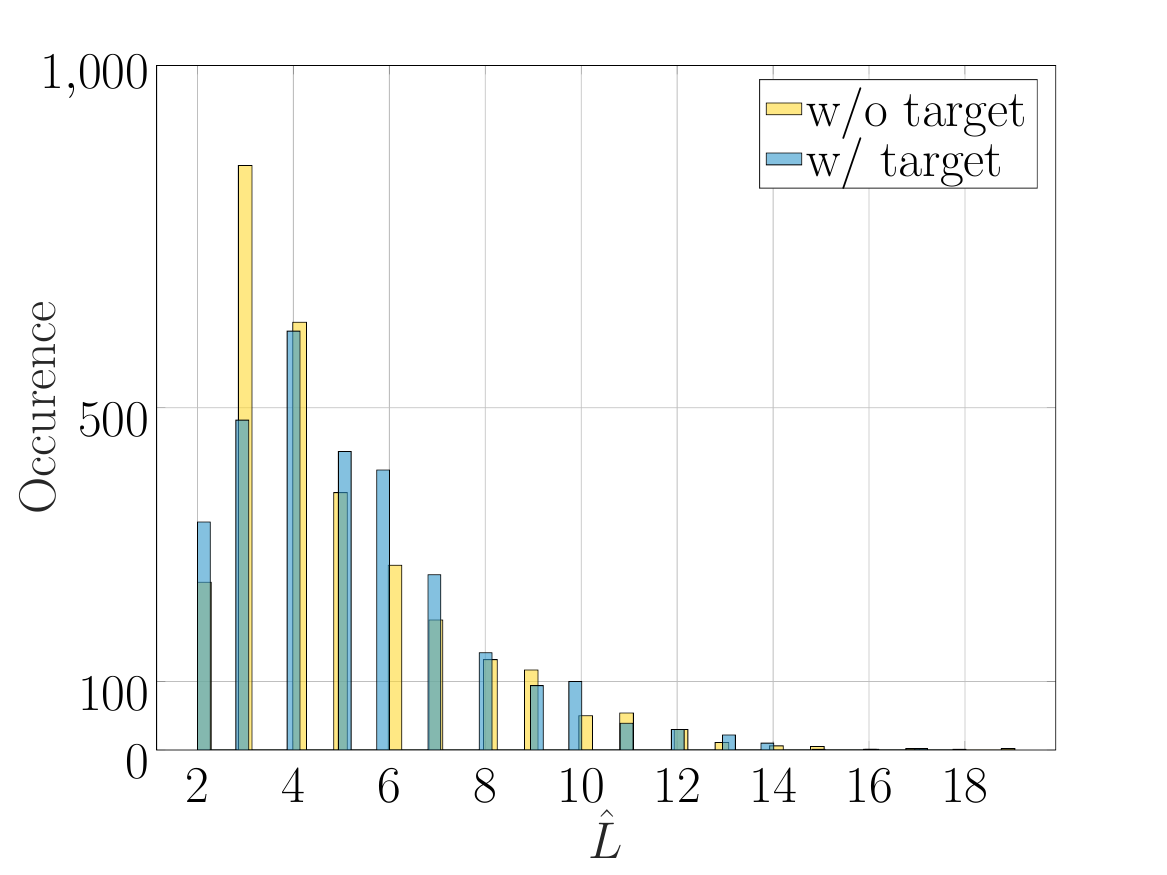}};
        \node[draw, fill=white, align=center] at (0.7,0.5) {\small Orientation: All};
    \end{tikzpicture}
    \caption{Frequency $\SI{6.5}{\GHz}$. Histograms of the estimated number of multipath components, with and without target.}
    \label{fig:hist-6}
\end{figure*}
\begin{figure*}[h]
    \centering
    \begin{tikzpicture}
        \node[inner sep=0] at (0,0) {\includegraphics[width=0.25\textwidth]{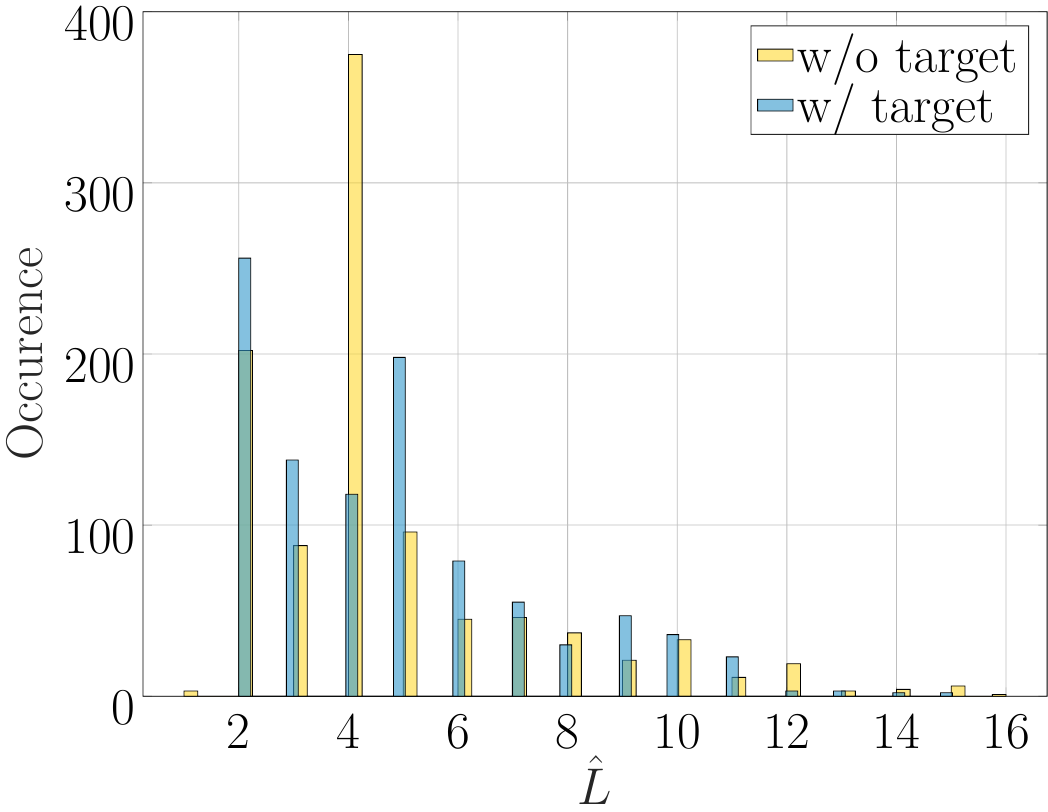}};
        \node[draw, fill=white, align=center] at (0.8,0.5) {\small Orientation: Beta};  
    \end{tikzpicture}
        \begin{tikzpicture}
        \node[inner sep=0] at (0,0) {\includegraphics[width=0.25\textwidth]{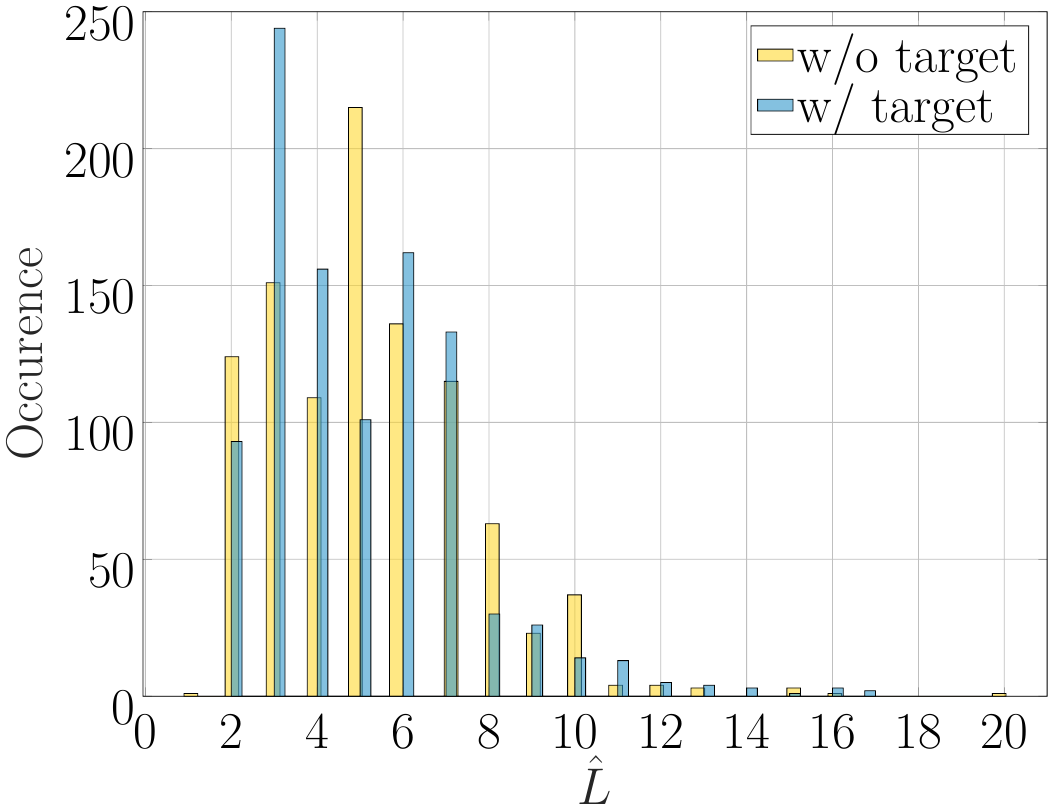}};
        \node[draw, fill=white, align=center] at (0.85,0.77) {\small Orientation: Alpha};
    \end{tikzpicture}
        \begin{tikzpicture}
        \node[inner sep=0] at (0,0) {\includegraphics[width=0.25\textwidth]{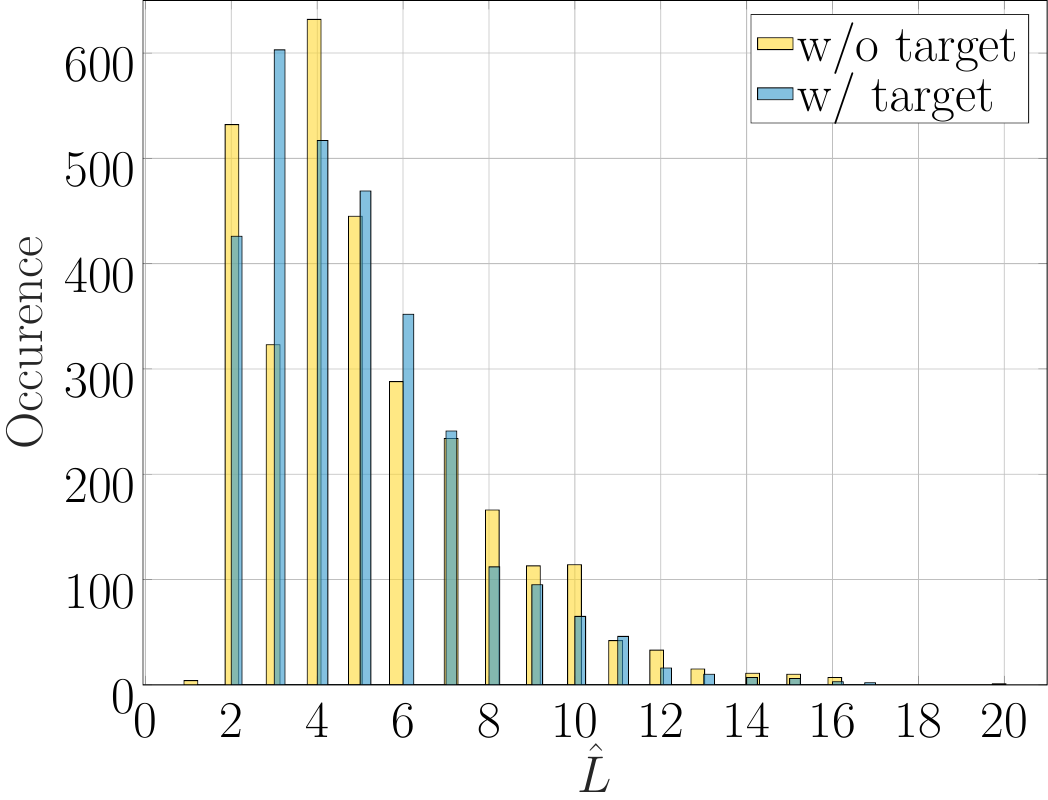}};
        \node[draw, fill=white, align=center] at (0.8,0.5) {\small Orientation: All};
    \end{tikzpicture}
    \caption{Frequency $\SI{8.75}{\GHz}$. Histograms of the estimated number of multipath components, with and without target.}
    \label{fig:hist-8}
\end{figure*}

\begin{figure}[!t]
\centering
\includegraphics[width=2.25in]{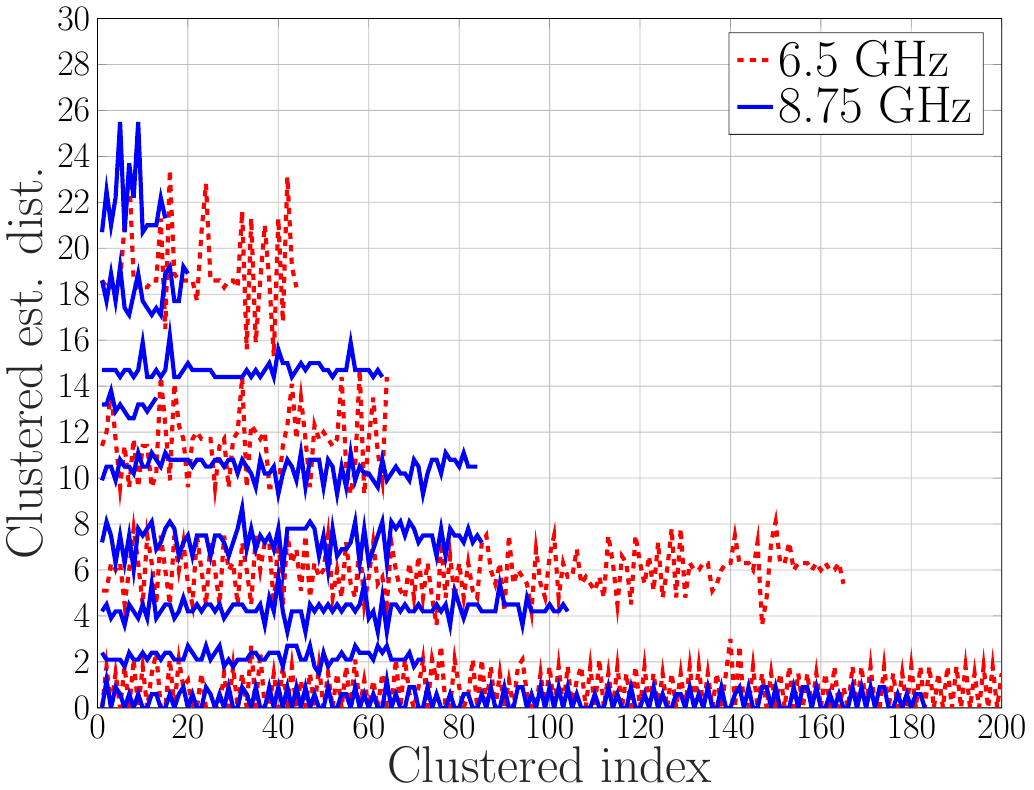}
\caption{The $K-$means clustered paths for $\SI{6.5}{\GHz}$ and  $\SI{8.75}{\GHz}$ showing path distances as a function of their index.}
\label{fig:clustered}
\end{figure}

\subsection{Comparison between CIR and MUSIC-PDP}
In Fig. \ref{fig:PDP_all}, we analyze the instantaneous MUSIC-PDP after frequency smoothing, whilst comparing it with the \ac{CIR}-\ac{PDP} obtained from \eqref{eq:h_hat}. 
Clearly, the frequency-smoothed \ac{MUSIC}-\ac{PDP} provides a refined estimation of the multipath channel via its orthogonality features, given a good model order selection, e.g. the elbow method.
It is worth noting that the estimator can in some cases capture an estimate multipath with a delay closely matching a path reflected by the target, as can be seen in Fig.~\ref{fig:PDP_all}. 
Also, when the target is placed, a new path is created at around $\SI{4.7}{\metre}$ which is a result of multiple bounces in the indoor environment.

In the following subsection, we provide a statistical analysis of how the number of multipath changes with and without the target for different frequencies.

\subsection{Multi-band Analysis}
In Fig. \ref{fig:hist-6}, we present histograms of the estimated number of multipath components at $\SI{6.5}{\GHz}$, comparing scenarios with and without a target for various orientations. The case of $\SI{8.75}{\GHz}$ is shown in Fig. \ref{fig:hist-8}.
The histogram was computed by estimating the number of multipath components over all positions and antennas. 

For orientation Beta and frequency $\SI{6.5}{\GHz}$, Fig. \ref{fig:hist-6}, we see that the mode of the histogram occurs at $4$ when no target is present as compared to $5$ when the target is present. 
A different behavior occurs at $\SI{8.75}{\GHz}$, Fig. \ref{fig:hist-8}, where a mode without a target is $4$ as compared to $2$ when a target is present. 
For orientation Alpha in the $\SI{6.5}{\GHz}$ band, the histogram mode is at $3$ when no target is present, increasing to $4$ with the target reflector. 
Interestingly, at a higher frequency of $\SI{8.75}{\GHz}$, the mode shifts to $5$ without a target and decreases to $3$ when a target is present.
Averaging across all orientations (namely Alpha, Beta, and Gamma), Fig. \ref{fig:hist-6} shows the histogram mode at $3$ without a target, which rises to $4$ when the target reflector is introduced. Notably, at the higher frequency of $\SI{8.75}{\GHz}$, illustrated in Fig. \ref{fig:hist-8}, the mode changes to $4$ in the absence of a target and drops to $3$ when a target is present.

These observations highlight the frequency-dependent nature of multipath components in the presence of a target reflector. 
At lower frequencies, such as $\SI{6.5}{\GHz}$, longer wavelengths tend to diffract around obstacles, maintaining or even increasing the number of observed multipath components when a target is introduced, as evidenced by shifts in histogram modes across orientations. 
Higher frequencies, like $\SI{8.75}{\GHz}$, tend to behave differently.
Due to their shorter wavelengths, they are more susceptible to blockage and reflection effects, which can reduce the multipath component density when a target is present. 
This phenomenon is particularly apparent in orientation Beta, where the histogram mode decreases from $4$ to $2$ with the target. 

\begin{figure*}
    \centering
    \includegraphics[width=.9\linewidth]{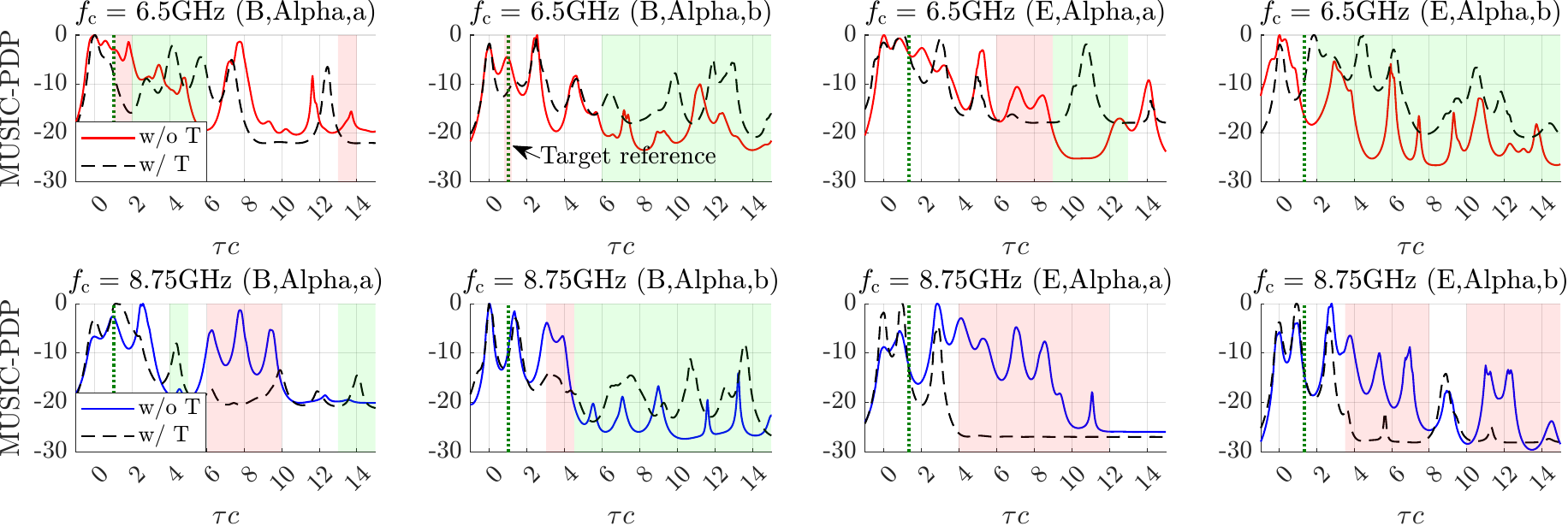}
    \caption{MUSIC-PDP with \textit{P}-region (green shaded) and \textit{N}-region (red shaded), with and without target (T).}
    \label{fig:music-pdp}
\end{figure*}

In Fig. \ref{fig:clustered}, we plot the $K$-means output where each line shows the delay estimates, translated into distances, as a function of the clustering index.
The clustering index is used to index the delay estimates belonging to a single cluster.
It appears that $K$-means clustering can classify more stable paths at the higher frequency, i.e. at $\SI{8.75}{\GHz}$, as compared to the lower frequencies.
Moreover, there appear to be paths that fluctuate for both frequencies, after clustering, with more fluctuations at lower frequencies. We can also see that $K$-means can distinguish more paths at higher frequencies than those at lower ones.
This can also indicate that \textit{higher frequencies can promote more distinguishable paths}.

\subsection{Insights for Multi-Band Sensing}
In the following, we provide an illustration of the data to better understand the results of Figs.~\ref{fig:hist-6}, \ref{fig:hist-8}, and \ref{fig:clustered}, which is also an insightful analysis of the impact of the frequency on ISAC. The delay axis is given is converted to traveled by the path.
Our approach is to provide a brief illustration of how the energy distribution on the channel impulse changes with the presence of the target in some cases.
We define two energy regions on the channel impulse response, namely, \textit{positive (P)} and \textit{negative (N)}.

\begin{itemize}
   \item \textit{P}-region: parts of the channel where new reflections are created. These regions are particularly interesting for bi-static sensing because the new paths do not interfere with the clutter in the time domain, facilitating sensing. 
    While typical bistatic sensing models rely on a single bounce from the target to the receiver, the \textit{P}-region can contain late paths that also convey information about the target.
    \item \textit{N}-region: parts of the channel where paths are blocked.
    Another approach to perform sensing is to analyze the multipath components that are suppressed in the presence of the target. When prior knowledge of the clutter is available, this can be an invaluable source of information to infer the target's spatial characteristics.
\end{itemize}

The results are shown in Fig.~\ref{fig:music-pdp}. As predicted by the histograms of Fig.~\ref{fig:hist-6}, we observe that the $\SI{6.5}{\GHz}$ frequency has a tendency to have newer paths with the target (\textit{P}-region with green shaded areas), while the $\SI{8.75}{\GHz}$ has more blockage (\textit{N}-region with red shaded areas).
In general, these results indicate that \textit{one effective way to perform sensing in high frequencies is by analyzing the clutter blockage, while lower frequencies create late paths (multi-bounce) with higher energy.}

\section{Conclusion}\label{sec:conclusion}
In this paper, we conducted a set of indoor channel measurements for the low FR3 frequencies $\SI{6.5}{\GHz}$ and $\SI{8.75}{\GHz}$, using the Pi-Radio FR3 front-end board which allows generating RF signals anywhere between $\SI{6}{\GHz}$ and $\SI{24}{\GHz}$.
In addition to multi-band, we also measured the channel in the presence of a target to investigate sensing aspects.
To analyze the data, we have used an established post-processing methodology based on MUSIC, which provides a refined estimation of the channel impulse response delays. The analysis revealed that in the presence of the target, the $\SI{6.5}{\GHz}$ has a tendency to add new distinguishable reflections, while the higher frequency $\SI{8.75}{\GHz}$ is more susceptible to blockage.

The methodology presented in this paper to measure sensing and communication aspects of the channel in a multi-band setting will serve as a guideline for future work. In particular, we plan to extend this analysis by characterizing spatial-frequency features across the entire FR3 spectrum.

\bibliography{references}
\bibliographystyle{IEEEtran}

\vfill

\end{document}